\documentclass[11pt,twocolumn]{article}
\usepackage{amssymb,epsfig,wrapfig,graphics,aas_macros,appendix}
\usepackage[dvips]{color}
\usepackage[footnotesize]{caption}

\def\comment#1{{}}

\newlength{\cvindent}
\newlength{\cvhang}

\newlength{\refindent}
\begin{document}

\vspace{-0.5in}

\title{Gamma Ray Burst Section of the White Paper on the Status and Future of ground-based TeV gamma-ray astronomy \\[18pt] 
\large{A.~D.~Falcone$^{1,2}$,
D.~A.~Williams$^3$, 
M.~G.~Baring$^4$, 
R.~Blandford$^5$, 
J.~Buckley$^6$,
V.~Connaughton$^7$, 
P.~Coppi$^8$, 
C.~Dermer$^9$, 
B.~Dingus$^{10}$, 
C.~Fryer$^{10}$,
N.~Gehrels$^{11}$,
J.~Granot$^5$, 
D.~Horan$^{12,13}$, 
J.~I.~Katz$^6$, 
K.~Kuehn$^{14}$, 
P.~M\'esz\'aros$^{1,2}$, 
J.~Norris$^{15}$,
P.~Saz~Parkinson$^3$, 
A.~Pe'er$^{16}$, 
E.~Ramirez-Ruiz$^{17}$,
S.~Razzaque$^9$, 
X.~Y.~Wang$^{18}$,
B.~Zhang$^{19}$}}
\date{}
\maketitle

\footnotetext[1] {Department of Astronomy \& Astrophysics, Pennsylvania State University, University Park, PA 16802, USA}
\footnotetext[2] {Center for Particle Astrophysics, Pennsylvania State University, University Park, PA 16802, USA}
\footnotetext[3] {Santa Cruz Institute for Particle Physics, University of California, Santa Cruz, CA 95064, USA}
\footnotetext[4] {Department of Physics and Astronomy, Rice University, Houston, Texas 77251-1892, USA}
\footnotetext[5] {KIPAC, Stanford Linear Accelerator Center, Stanford University, Stanford, CA 94309, USA}
\footnotetext[6] {Department of Physics, Washington University in St Louis, St Louis, MO 63130, USA}
\footnotetext[7] {National Space Science and Technology Center, Huntsville, AL 35805, USA}
\footnotetext[8] {Department of Astronomy, Yale University, New Haven, CT 06520-8101, USA}
\footnotetext[9] {US Naval Research Laboratory, 4555 Overlook SW, Washington, DC, 20375-5352, USA}
\footnotetext[10] {Los Alamos National Lab, Los Alamos, NM 87545, USA}
\footnotetext[11] {NASA/Goddard Space Flight Center, Greenbelt, MD 20771, USA}
\footnotetext[12] {Argonne National Laboratory, 9700 South Cass Avenue, Argonne, IL 60439, USA}
\footnotetext[13] {now at Laboratoire Leprince-Ringuet, Ecole Polytechnique, CNRS/IN2P3, Palaiseau, France}
\footnotetext[14] {Department of Physics, Ohio State University, Columbus, OH 43210, USA}
\footnotetext[15] {University of Denver, Denver CO 80208, USA}
\footnotetext[16] {Space Science Telescope Institute, Baltimore, MD 21218, USA}
\footnotetext[17] {Department of Astronomy, University of California, Santa Cruz, CA 95064, USA}
\footnotetext[18] {Department of Astronomy, Nanjing University, Nanjing 210093, China}
\footnotetext[19] {Department of Physics and Astronomy, University of Nevada, Las Vegas, NV 89154-4002, USA}

\section{Introduction}


High energy astrophysics is a young and relatively undeveloped
field, which owns much of the unexplored ``discovery space'' in
contemporary astronomy. The edge of this discovery space has recently been illuminated by the current generation of very high energy (VHE) telescopes, which have discovered a diverse catalog of more than seventy VHE sources. At this time, gamma ray bursts (GRBs) have eluded attempts to detect them with VHE telescopes (although some tentative, low-significance detections have been reported). However, theoretical predictions place them 
near the sensitivity limits of current instruments.
The time is therefore at hand to increase VHE telescope sensitivity, thus facilitating the detection of these extreme and mysterious objects. 


Much has been learned since the discovery of GRBs in the late 1960s.
There are at least two classes of GRB, most conveniently referred to as ``long'' and ``short,''
based on the duration and spectral hardness of their prompt sub-MeV emission.   The
distribution of the types and star formation rates of the host galaxies suggests different 
progenitors for these two classes.
The exact nature of the progenitors 
nevertheless 
remains unknown, although it is widely believed that long GRBs come from the deaths of massive rotating stars and short GRBs result from compact object mergers. The unambiguous solution to this mystery is critical to astrophysics since it has fundamental importance to several topics, including stellar formation history and ultra high energy cosmic ray acceleration. A detection of VHE emission from GRBs would severely constrain the physical parameters surrounding the particle acceleration from GRBs and the energy injected into the particle acceleration sites, and would therefore constrain the properties of the GRB progenitors themselves. These same observations would constrain models for cosmic ray acceleration.

One of the big questions regarding GRBs is 
whether the jets are dominated by ultrarelativistic protons, that interact with either the radiation field or the
background plasma, or 
are dominated by e$^+$e$^-$ pairs.  
The combination of Fermi and
current generation VHE 
telescopes 
such as 
HESS,
MAGIC  
and VERITAS
will 
contribute to progress on these questions in the near term, but more sensitive observations will likely be needed.

The same shocks which are thought to accelerate electrons
responsible for non-thermal $\gamma$-rays in GRBs should also
accelerate protons. Both the internal and the external reverse shocks
are expected to be mildly relativistic, and are expected to lead to relativistic
protons. The maximum proton energies achievable in GRB shocks are estimated to be 
$\sim$10$^{20}$ eV, comparable to the highest energies of the mysterious ultra high energy cosmic rays measured with large ground arrays. 
The accelerated protons can interact with the fireball photons,
leading to 
pions, followed by high-energy gamma rays, 
muons, and neutrinos. Photopion production
is enhanced in conditions of high internal photon target density, 
whereas 
if the density of (higher-energy) photons is too large,
the fireball is optically thick to gamma-rays, 
even in a purely leptonic outflow.
High-energy gamma-ray studies of GRBs 
provide a direct probe of the shock
proton acceleration as well as of the photon density.

\subsection{Status of theory on emission\\models}

Gamma-ray burst $\nu F_\nu$ spectra have a peak at photon energies ranging from a few keV to several MeV, and the spectra are nonthermal. From EGRET data, it is clear that the 
spectra extend to at least several GeV \cite{Hurleyetal94,Dingus95,Dingusetal98,Gonzalezetal03}, and there is a possible detection in the TeV range by Milagrito \cite{Atkinsetal00,Atkinsetal03}.
These non-thermal spectra imply
that a significant fraction of the explosion energy is first converted into another form of energy before being dissipated and converted to nonthermal radiation. The most widely accepted interpretation is the conversion of the explosion energy into kinetic energy of a
relativistic flow \cite{Paczynski86,Goodman86,Paczynski90}. At a second stage,
the kinetic energy is converted into radiation via internal collisions 
(internal shock model) resulting from variability
in the ejection from the progenitor \cite{PaczynskiXu94,ReesMeszaros94} or an
external collision (external shock model) with the surrounding medium 
\cite{ReesMeszaros92,DermerMitman99,Dermer07}. The collisions
produce shock waves, which enhance and are believed to create magnetic fields, as well as to accelerate electrons to high energies \cite{Kazimuraetal98,Silvaetal03,Frederiksenetal04,%
Nishikawaetal05,Spitkovsky08}. 
In the standard theoretical model, the initial burst of emission described above (prompt emission) is followed by 
afterglow emission, discussed below, from an external shock that moves through the circumburst environment.

Flux variability in GRBs is seen on timescales as short as milliseconds and can occur at late times.
This rapid variability 
can be easily explained in the internal shock model, which
makes it the most widely used model. It can also be explained in the context of the external
shock model 
either 
if one assumes variations in the strength of the magnetic field or in the energy
transfer to the non-thermal electrons \cite{PanaitescuMeszaros98}, or 
by collisions of the outflow with  
small, high density clouds in the surrounding medium \cite{DermerMitman99,Dermer07}.

An alternative way of producing the emission involves conversion of the explosion energy
into magnetic energy \cite{Thompson94,Usov94,MeszarosRees97}, which
produces a flow that is Poynting-flux dominated. The emission is produced following dissipation
of the magnetic energy via reconnection of the magnetic field lines \cite{Drenkhahn02,%
DrenkhahnSpruit02,LyutikovBlandford03,GianniosSpruit05}. 
An apparent advantage of this model over the internal or external shock model is that the conversion of
energy to radiation is much more efficient (see \cite{Kumar99,Panaitescuetal99} on the
efficiency problem in the internal shocks model). The microphysics of the
reconnection process in this model, like the microphysics determining the 
fraction of energy in relativistic electrons and in the magnetic field
in the internal and external 
shock scenarios,
is not yet fully understood.

VHE observations 
probe the extremes of the efficiency of energy conversion for each of these models 
and 
simultaneously probe the environment where the emission originated. 


The dissipation of kinetic and/or magnetic energy leads to the emission of radiation.
The leading emission mechanism employed to interpret the GRB prompt emission in the keV-MeV region of the spectral energy distribution is nonthermal synchrotron radiation \cite{Meszarosetal93,MeszarosRees93b,Katz94b}.
An order of magnitude estimate of the maximum observed energy of photons produced
by synchrotron emission was derived in \cite{PeerWaxman04b}: Assuming that the
electrons are Fermi accelerated in the shock waves, the maximum Lorentz factor of the accelerated
electrons $\gamma_{\max}$ is found by equating the particle acceleration time
and
the synchrotron cooling time,
yielding $\gamma_{\max} = 10^5 / \sqrt{B /10^6}$, where $B$ is the magnetic
field strength in gauss.
For relativistic motion with bulk Lorentz factor $\Gamma$ 
at redshift $z$, synchrotron emission from electrons with $\gamma_{\max}$ 
peaks in the observer's frame at
energy  $70 \, (\Gamma/315) 
(1+z)^{-1}$ GeV,  
which is independent of the magnetic field.
Thus, synchrotron emission can 
produce photons 
with energies up to, and possibly exceeding, $\sim$100 GeV.

Many of the observed GRB spectra were found to be consistent with the synchrotron
emission interpretation \cite{Tavani96a,Tavani96b,Cohenetal97}. However, a significant fraction
of the observed spectra were found to be too hard (
spectral photon index harder than $2/3$ at low energies) to be accounted for by this model \cite{Crideretal97,Preeceetal98,Fronteraetal00,Preeceetal02,Ghirlandaetal03}.
This motivated studies of magnetic field tangling on very short
spatial scales \cite{Medvedev00}, anisotropies in the electron 
pitch angle distributions \cite{LloydPetrosian00,LloydRonningPetrosian02}, 
reprocessing of radiation by an optically thick cloud heated by the impinging gamma rays \cite{DermerBoettcher00} or by synchrotron self absorption \cite{Granotetal00}, 
and
the contribution of a photospheric (thermal) component \cite{MeszarosRees00,DaigneMochkovitch02,Meszarosetal02}. 
A
thermal component that accompanies the first stages of the overall non-thermal emission and decays after a few seconds was consistent with some observations \cite{Ryde04,Ryde05}. Besides explaining the hard spectra observed in some of the 
GRBs seen by the Burst and Transient Source Experiment (BATSE),
the thermal component provides seed photons that can be Compton scattered by relativistic electrons, resulting in a potential VHE gamma ray emission signature that can be tested.



A natural emission mechanism that can contribute to emission at high energies 
($\gtrsim$MeV) is inverse-Compton (IC) scattering. The seed photons for the scattering can be synchrotron photons emitted
by the same electrons, namely synchrotron self-Compton (SSC) emission 
\cite{MeszarosRees94,Meszarosetal94,PapathanassiouMeszaros96,%
Liangetal97,SariPiran97,PillaLoeb98,ChiangDermer99,PanaitescuMeszaros00},
although in some situations this generates MeV-band peaks broader than those
observed \cite{BaringBraby04}.
The seed photons can also be  thermal emission originating from the photosphere 
\cite{ReesMeszaros05,Peeretal06}, 
an accretion disk 
\cite{ShavivDar95}, 
an accompanying supernova 
remnant \cite{Lazzatietal00,Broderick05}, or supernova emissions in two-step collapse scenarios \cite{Inoueetal03}.
Compton scattering of photons can produce emission up to observed energies
$15 \, (\gamma_{\max}/10^5)\,(\Gamma/315) 
(1+z)^{-1}$ TeV, 
well into the VHE regime.

The shapes of the Comptonized emission spectra in GRBs depend on the spectra of the seed photons and the energy and pitch-angle distributions of the electrons. A thermal population of electrons can inverse-Compton scatter seed thermal photons \cite{Liang97} or photons at energies below the synchrotron self-absorption frequency to produce the observed peak at sub-MeV energies \cite{GhiselliniCelotti99}. Since the electrons cool by the IC process, a variety of spectra can be obtained \cite{PeerWaxman04b,Peeretal06}.
Comptonization can produce a dominant high-energy component \cite{GuettaGranot03a}
that can explain hard high-energy spectral components, such as that observed in
GRB 941017 \cite{Gonzalezetal03,GranotGuetta03a,PeerWaxman04a}. 
Prolonged higher energy emission could potentially be observed with a sensitive VHE gamma ray instrument.


The maximum observed photon energy
from GRBs 
is limited by the annihilation of  gamma rays with target photons, both extragalactic IR background and photons local to the GRB, to produce electron-positron pairs. This limit is sensitive to the uncertain value of the bulk motion Lorentz factor as well as to the spectrum at low energies, and is typically in the sub-TeV regime.
Generally, escape of high-energy photons requires large Lorentz factors.
In fact, observations of GeV photons have been used to constrain the minimum Lorentz factor of the bulk motion of the flow \cite{KrolikPier91,FenimoreEpsteinHo93,%
WoodsLoeb95,BaringHarding97,LithwickSari01,Razzaqueetal04}, 
and spectral 
coverage up to TeV energies could 
further constrain
the Lorentz factor \cite{KobayashiZhang03,Zhangetal03,Peeretal07}.
If the Lorentz factor can be determined independently, e.g.\ from afterglow modeling,
then the annihilation signature can be used to diagnose the gamma-ray emission 
region \cite{GuptaZhang08}.
The evidence for acceleration of leptons in GRB blast waves is based on fitting lepton synchrotron spectra models to GRB spectra. This consistency of leptonic models with observed spectra still allows the possibility of hadronic components in these bursts, and perhaps more importantly, GRBs with higher energy 
emission
have not been explored for such hadronic components due to the lack of sensitive instruments in the GeV/TeV energy range. The crucially important high-energy emission components, represented by only 5 EGRET spark chamber bursts, a handful of BATSE and EGRET/TASC GRBs, and a marginal significance Milagrito TeV detection, were statistically inadequate to look for 
correlations between high-energy and keV/MeV emission
that can be attributed to a particular 
process. Indeed, the 
prolonged
high-energy components in GRB 940217 and the ``superbowl'' burst, GRB 930131, and the anomalous gamma-ray emission component in GRB 941017, behave quite differently than the measured low-energy gamma-ray light curves. Therefore, it is quite plausible that hadronic emission components are found in the high energy spectra of GRBs.

Several theoretical mechanisms exist for hadronic VHE emission components. 
Accelerated protons can emit synchrotron radiation in the GeV--TeV energy band \cite{Vietri97,BottcherDermer98,Totani98}. The power emitted by a particle is 
$\propto \gamma^2/m^2$, where $\gamma$ is the Lorentz factor of the particle and $m$ is its mass. 
Given the larger mass of the proton, 
to achieve the same output luminosity, 
the protons have 
$\sim$1836 times 
higher mean Lorentz factor, 
the acceleration mechanism must convert 
$\sim$ 3 million times 
more energy to protons than electrons 
and the peak of the proton emission would be at 
$\gtrsim$ 2000 times higher  
energy 
than the peak energy of photons emitted by the electrons. 
Alternatively, 
high-energy baryons can produce energetic pions, via photomeson interactions with the low energy photons, creating 
high-energy photons and neutrinos following the pion decay \cite{BottcherDermer98,WaxmanBahcall97,WaxmanBahcall00,DermerAtoyan03,GuettaGranot03b,GranotGuetta03b}. This process could be the primary source of ultra high energy (UHE)
neutrinos. 
Correlations between gamma-ray opacity, bulk Lorentz factor, and neutrino production will test whether GRBs are UHE cosmic ray sources \cite{Dermeretal07}. 
If the neutrino production is too weak to be detected, then the former two measurements can be obtained independently with sensitive GeV-TeV $\gamma$-ray telescopes and combined to test for UHE cosmic ray production. 
Finally, 
proton-proton or proton-neutron collisions may also be 
a
source of 
pions 
\cite{PaczynskiXu94,Katz94a,DePaolisetal00,Derishevetal99,BahcallMeszaros00}, and in addition, 
if there are neutrons in the flow, then the neutron $\beta$-decay has a drag effect on the protons, which may produce another source of radiation \cite{Rossietal06}. Each of 
these cases has a VHE 
spectral shape and intensity 
that can be studied coupled with the 
emission measured at lower energies and with neutrino measurements.

Afterglow emission is explained in synchrotron-shock models by the same processes that occur during the prompt phase. The key difference is that the afterglow emission originates from large radii,
$\gtrsim 10^{17}$~cm, as opposed to the much smaller radius of the flow during the prompt emission
phase,  $\simeq 10^{12} - 10^{14}$~cm for internal shocks, and $\simeq 10^{14} - 10^{16}$~cm for external shocks. As a result, the density of the blast-wave shell material is smaller during the afterglow emission phase than in the prompt phase, and some of the radiative mechanisms, e.g. thermal collision processes, may become less important.


Breaks in the observed lightcurves,
abrupt changes in the power law slope, 
are attributed to a variety of phenomena, such as refreshed shocks originating from late time central engine activity \cite{ReesMeszaros98,Granotetal03}, aspherical variations in the energy \cite{KumarPiran00}, or variations in the external density \cite{WangLoeb00,Nakaretal03}. Blast wave energy escaping in the form of UHE neutrals and cosmic rays can also produce a rapid decay in the X-ray light curve \cite{Dermer07}. In addition, interaction of the blast wave with the wind termination shock 
of the progenitor 
may be the source of a jump in the lightcurve \cite{Wijers01,RamirezRuizetal05,PeerWijers06}, 
although this bump may not be present at a significant level \cite{NakarGranot07}. High-energy gamma-ray observations may show whether new photohadronic emission mechanisms are required, or if the breaks do not require new radiation mechanisms for explanation (see, e.g., \cite{Genetetal07,UhmBeloborodov07}).

\subsection{GRB Progenitors}

We still do not know the exact progenitors of GRBs, and it is therefore difficult, if not impossible, to understand the cause of these cosmic explosions. These GRB sources involve emission of energies that can
exceed $10^{50}$ ergs. The seat of this activity is
extraordinarily compact, as indicated by rapid variability of the
radiation flux on time scales as short as milliseconds. It is unlikely
that mass can be converted into energy with better than a few (up to
ten) percent efficiency; therefore, the more powerful short GRB sources
must ``process'' upwards of $10^{-3}M_\odot$ through a region which is
not much larger than the size of a neutron star (NS) or a stellar mass
black hole (BH). No other entity can convert mass to energy with such
a high efficiency, or within such a small volume.  
The leading contender for the production of the longer class of GRBs
---  supported by observations of supernovae associated with several bursts --- 
is the catastrophic collapse of massive, rapidly rotating stars.
The current preferred model for short bursts, the merger of binary systems of
compact objects, such as double neutron star systems (e.g. Hulse-Taylor pulsar systems)
is less well established. 
A fundamental problem posed by GRB sources is how to generate
over $ 10^{50}$ erg in the burst nucleus and channel it into
collimated relativistic plasma jets.

The progenitors of GRBs are essentially masked by the resulting fireball, which reveals little more than
the basic energetics and microphysical parameters of relativistic shocks.  
Although long and short bursts most likely have different progenitors, the observed radiation
is very similar.
Progress in understanding the progenitors can come from determining the burst environment, the kinetic energy and Lorentz factor of the ejecta, the duration of the central engine activity, and the redshift distribution. VHE gamma-ray observations can play a supporting role in this work.  To the extent that we understand GRB emission across the electromagnetic spectrum, we can look for the imprint of the burst environment or absorption by the extragalactic background light on the spectrum as an indirect probe of the environment and distance, respectively. VHE 
emission
may also prove to be crucial to the energy budget of many bursts, thus constraining the progenitor.

\section{High-energy observations of gamma-ray bursts}

Some of the most significant advances in GRB research have come from
GRB correlative observations at longer wavelengths. Data on
correlative observations at shorter wavelengths are sparse but
tantalizing and inherently very important. One definitive observation
of the prompt or afterglow emission could significantly influence our
understanding of the processes at work in GRB emission and its
aftermath. Although many authors have predicted its existence, the predictions
are near or below the sensitivity of current instruments, and there
has been no definitive detection of VHE emission from a GRB either
during the prompt phase or at any time during the multi-component
afterglow.

For the observation of photons of energies above 300\,GeV, only
ground-based telescopes are available. These ground-based
telescopes fall into two broad categories, air shower arrays and
imaging atmospheric Cherenkov telescopes (IACTs). The air shower arrays,
which have wide fields of view that are suitable 
for GRB searches, are relatively insensitive. There are several reports
from these instruments of possible TeV emission: 
emission $>$16 TeV from GRB\,920925c \cite{Padilla:98:Airobicc},
an indication of 10\,TeV emission in a stacked analysis of 57 bursts
\cite{Amenomori:01:TibetGRBs},
and
an excess gamma-ray signal during the prompt phase of
GRB\,970417a 
\cite{Atkinsetal00}. In all of
these cases however, the statistical significance of the detection is
not high enough to be conclusive. 
In addition to searching the Milagro
data for VHE counterparts for over 100 satellite-triggered GRBs since 2000
\cite{Atkins:05:MilagroGRBCounterparts,SazParkinson:07:MilagroGRBCounterparts,Abdo:07:MilagroGRBCounterparts},
the Milagro Collaboration
conducted a search for VHE transients of 40 seconds to 3 hours
duration in the northern sky \cite{Atkins:04:MilagroGRB}; 
no evidence
for VHE emission was found from 
these searches.

IACTs have better flux sensitivity and energy resolution 
than air shower arrays,
but are limited by their
small fields of view (3--5$^\circ$) and low duty cycle ($\sim$10\%).
In the BATSE \cite{Meegan:92} era (1991--2000), attempts at GRB monitoring were
limited by slew times and uncertainty in the GRB source position
\cite{Connaughton:97}. 
More recently, VHE upper
limits from 20\% to 62\% of the Crab flux at late
times ($\gtrsim$4 hours) were obtained with Whipple Telescope
for seven GRBs in 2002-2004 \cite{Horan:07}. 
The MAGIC Collaboration took
observations of GRB\,050713a beginning 40 seconds after the prompt
emission but saw no evidence for VHE emission
\cite{Albert:06:40sGRB}. Follow-up GRB observations have been made on
many more GRBs by the MAGIC Collaboration
\cite{Bastieri:07:ICRCStatus} but no detections have been
made \cite{Albert:2006:GRBs,Bastieri:7:ICRCGRBObs}.  Upper limits of 2--7\% of the 
Crab flux on the
VHE emission following three GRBs have also been obtained with VERITAS \cite{Horan:07ICRC}.

One of the main obstacles for VHE observations of GRBs is the distance
scale. Pair production interactions of gamma rays with the infrared
photons of the extragalactic background light attenuate the gamma-ray
signal,
limiting the distance over which VHE gamma rays can
propagate. 
The MAGIC
Collaboration 
has reported the detection of 
3C279, 
at 
redshift 
of 0.536
\cite{Teshima:07}. This represents a large increase in distance
to the 
furthest detected VHE source, revealing more of the
universe to be visible to VHE astronomers than was previously thought.






\section{High Energy Emission Predictions for Long Bursts}

As described earlier, long duration GRBs are generally believed to be associated with core
collapses of massive rotating stars \cite{Woosley93,MacFadyenWoosley99}, which lead to particle acceleration by relativistic internal shocks in jets. The 
isotropic-equivalent 
gamma-ray luminosity can vary
from $10^{47}~{\rm erg~s^{-1}}$ all the way to $10^{53}~{\rm erg~s^{-1}}$. 
They are distributed in a wide redshift range (from 
0.0085 for GRB 980425
\cite{Bloometal99} to 
6.29 for GRB 050904 \cite{Haislipetal06}, 
with a mean redshift of 
2.3--2.7
for Swift bursts, 
e.g. \cite{Bergeretal05a,Jakobssen06}). The low redshift long GRBs ($z \lesssim 0.1$, e.g. GRB 060218, $z=0.033$ \cite{Campanaetal06}) are typically sub-luminous with 
luminosities of 
$10^{47}-10^{49} ~{\rm erg~s^{-1}}$
and spectral peaks at lower energies,
so they are 
less likely 
detected at high energy.
However, one nearby, ``normal'' long GRB has been detected (GRB 030329, 
$z=0.168$), which has large fluences in both its prompt gamma-ray emission 
and afterglow.

\subsection{Prompt emission}
\begin{figure*}[!ht]
\centering
\includegraphics*[height=2in]{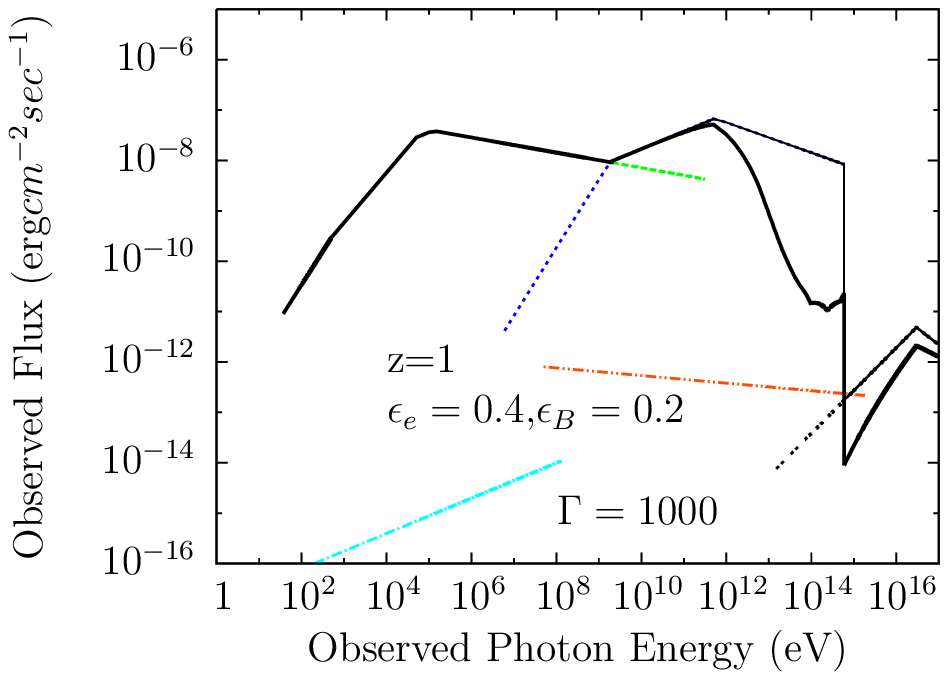}\quad
\includegraphics*[height=2in]{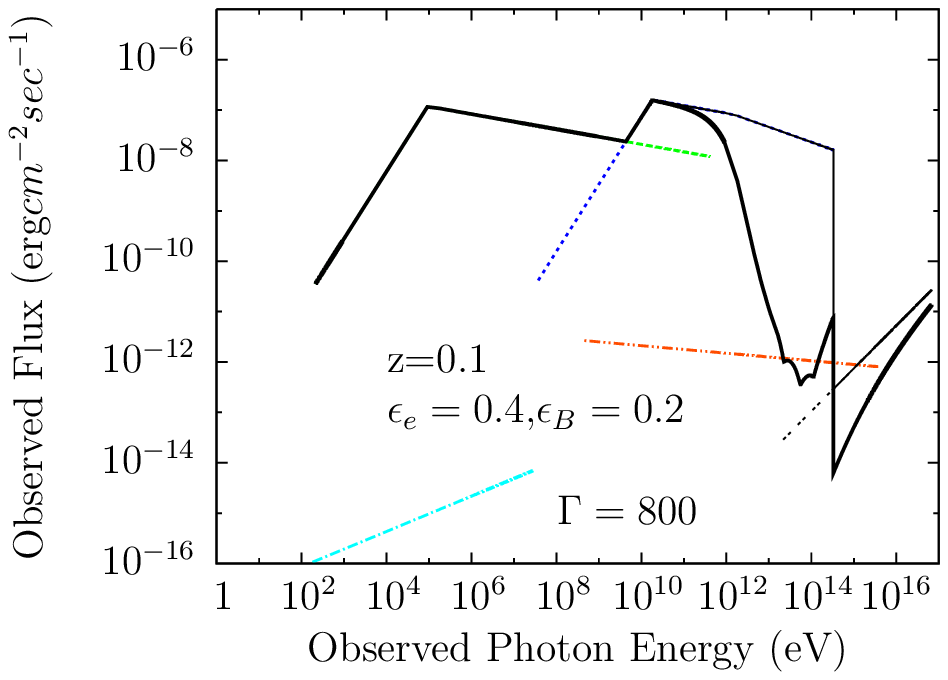}
\caption{Broad-band spectrum of the GRB prompt emission within the internal 
shock model 
(from \cite{GuptaZhang07}). 
(a) 
A 
long GRB with the observed sub-MeV luminosity of $\sim 10^{51}~{\rm
erg~s^{-1}}$,
is modeled 
for parameters as given in the figure.
The solid black lines represent the final spectrum before (thin line)
and after (thick line) including the effect of internal optical depths. 
The long dashed green line (mostly hidden) is the electron synchrotron component; the short-dashed blue line is the electron IC
component; the double short-dashed black curve on the right side is the $\pi^0$ decay 
component; the triple short-dashed dashed line represents the synchrotron radiation produced by 
e$^\pm$ from $\pi^\pm$ decays; the dash-dotted (light blue) line represents 
the proton synchrotron component. 
(b) The analogous spectrum of a bright short GRB with 
10$^{51}$ erg isotropic-equivalent energy release.} 
\label{low-compac}
\label{short}
\end{figure*}
The leading model of the GRB prompt emission is the internal shock model
\cite{ReesMeszaros94},
and we begin by discussing prompt emission in that context.
The relative importance of the leptonic vs. hadronic components for high
energy photon emission depends on the unknown shock equipartition parameters,
usually denoted as $\epsilon_e$, $\epsilon_B$ and $\epsilon_p$ for the
energy fractions carried by electrons, magnetic fields, and protons,
respectively. Since electrons are much more efficient emitters than protons,
the leptonic emission components usually dominate unless $\epsilon_e$ is 
very small. 
Figure~\ref{low-compac}a 
displays the broadband spectrum of a 
long GRB 
within the internal shock model for 
a particular choice
of parameters 
\cite{GuptaZhang07}. 
Since the phenomenological shock microphysics
is poorly known, modelers usually introduce 
$\epsilon_e$, $\epsilon_B$, $\epsilon_p$
as free parameters.
For 
$\epsilon_e$'s not too small (
$\gtrsim$10$^{-3}$), the high energy spectrum is dominated by the electron IC 
component, 
as 
in Fig.~\ref{low-compac}a.
For smaller $\epsilon_e$'s (e.g. $\epsilon_e =10^{-3}$), 
on the 
other hand, the hadronic components become at least comparable to the 
leptonic component above $\sim$100 GeV, and the $\pi^0$-decay component 
dominates the spectrum above $\sim$10 TeV. 

A bright GRB, 080319B, with a plethora of multiwavelength observations has recently allowed very detailed spectral modelling as a function of time, and it has shown that an additional high energy component may play an important role. For GRB 080319B, the bright optical flash suggests a synchroton origin for the optical emission and SSC production of the $\sim$500 keV gamma-rays \cite{Racusin08}.  The intensity of these gamma rays would be sufficient to produce a second-order IC peak around 10--100 GeV.

Due to the high photon number density in the emission region of GRBs, 
high energy photons 
have an optical depth for photon-photon pair production
greater than unity above a critical energy, producing a sharp spectral cutoff, which depends on the unknown bulk Lorentz 
factor of the fireball and the variability time scale of the central 
engine, which sets the size of the emission region.
Of course, the shape of time-integrated spectra will also be modified (probably to power laws rolling over to steeper power laws) due to averaging of evolving instantaneous spectra \cite{Granotetal08}. 
For the nominal bulk Lorentz factor $\Gamma=400$ (as suggested by recent 
afterglow observations, e.g. \cite{Molinarietal07}) and for a typical variability time 
scale 
$t_v = 0.01$ s,
the cut off energy is about several tens 
of GeV. Below 10 GeV, the spectrum is mostly dominated by the electron synchrotron
emission, so that with the observed high energy spectrum alone, usually there
is no clean 
differentiation of 
the leptonic vs. 
hadronic origin of the high energy gamma-rays. Such an issue may however be
addressed by collecting both prompt and afterglow data. Since a small $\epsilon_e$
is needed for a hadronic-component-dominated high energy emission, these fireballs
must have a very low efficiency 
for radiation,
$\lesssim\epsilon_e$,
and most
of the energy will be carried by the afterglow. As a result, a moderate-to-high 
radiative efficiency would suggest a leptonic origin of high energy photons, 
while a GRB with an extremely low radiative efficiency but an extended high 
energy emission component would be consistent with (but not a proof for)
the hadronic origin. If the fireball has a much larger Lorentz factor 
($\gtrsim 800$), 
the spectral cutoff energy 
is higher, as in Fig.~\ref{low-compac}. 
This would allow a larger spectral space to diagnose the origin of the GRB high energy emission and would place the cutoff energies in the spectral region that can only be addressed by VHE telescopes. 
At even higher energies, the fireball again becomes transparent to 
gamma rays 
\cite{Razzaqueetal04}, so that under ideal conditions, 
the $\sim$ PeV component due to $\pi^0$ decay
can escape the fireball. Emission above one TeV escaping from GRBs would suffer additional external 
attenuation by 
the cosmic infrared background (CIB) 
and the cosmic microwave background (CMB), thus limiting VHE observations of GRBs to lower redshifts (e.g. z$\lesssim$0.5--1).
The external shock origin of prompt emission is less favored by the Swift observations, which show a rapidly falling light curve 
following the prompt emission before the emergence of a more slowly
decaying component attributed to the external shock.
A small fraction of bursts lack the initial steep component,
in which case the prompt emission may result from an external
shock.  
Photons up to TeV energies are expected in the external
shock scenario \cite{Dermeretal00}, and
the 
internal pair cut-off energy
should be very high, 
more favorable to detection at VHE energies,
because of the less compact emission region.

The ``cannonball model'' of GRBs \cite{DardeRujula04}, in which the prompt GRB emission is produced by IC scattering from blobs of relativistic material (``cannonballs''), can also be used to explain the keV/MeV prompt emission, but it does not predict significant VHE emission during the prompt phase. Sensitive VHE observations would provide a strong constraint to differentiate between these models. The cannonball model could still produce delayed VHE emission during the deceleration phase, in much the  same way as the fireball model: as a consequence of IC scattering from relativistic electrons accelerated by the ejecta associated with the burst \cite{DadoDar05}.

\subsection{Deceleration phase}
\begin{figure*}[!ht]
\centering
\includegraphics[height=2.25in]{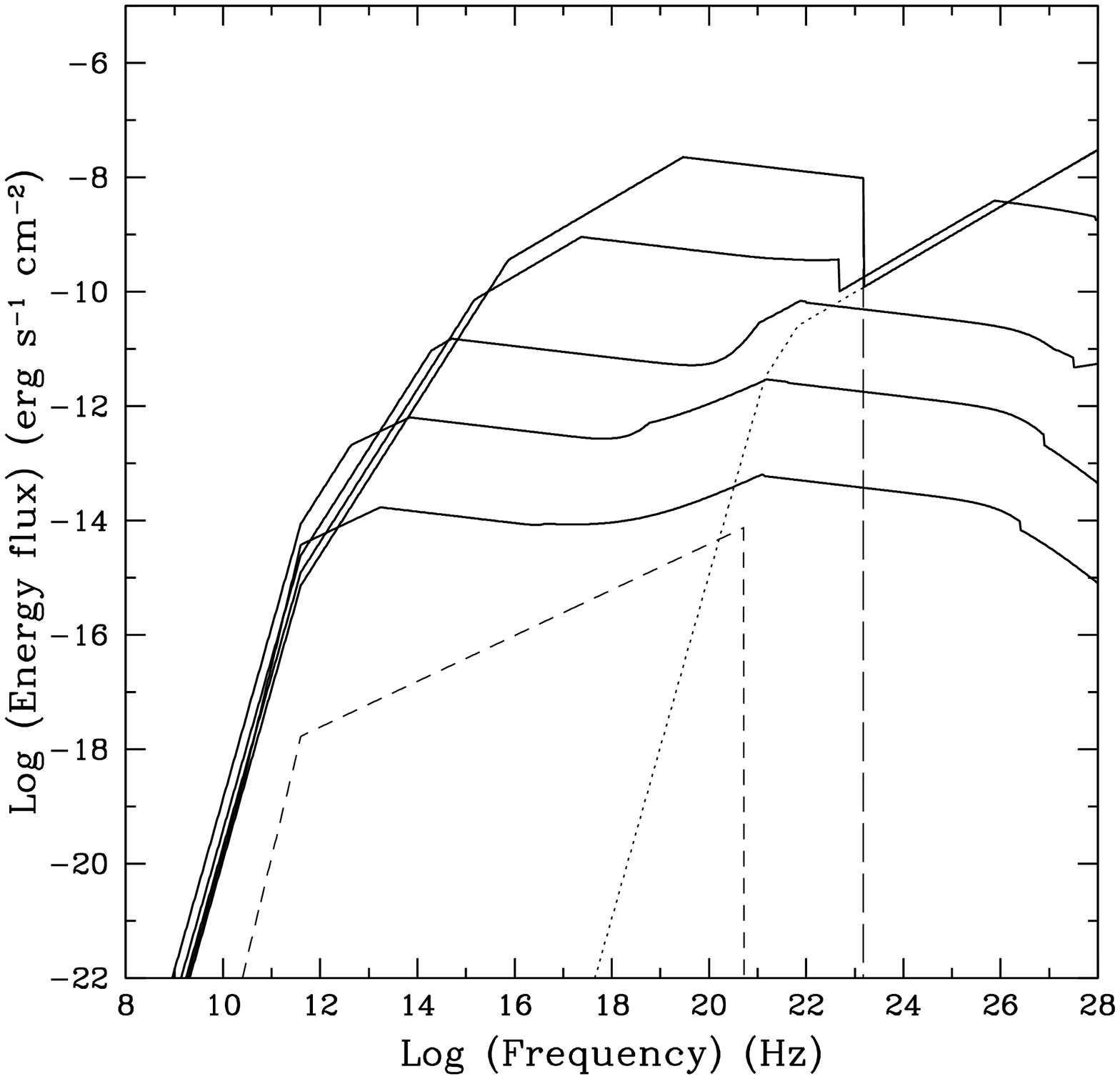} \quad
\includegraphics[height=2.25in]{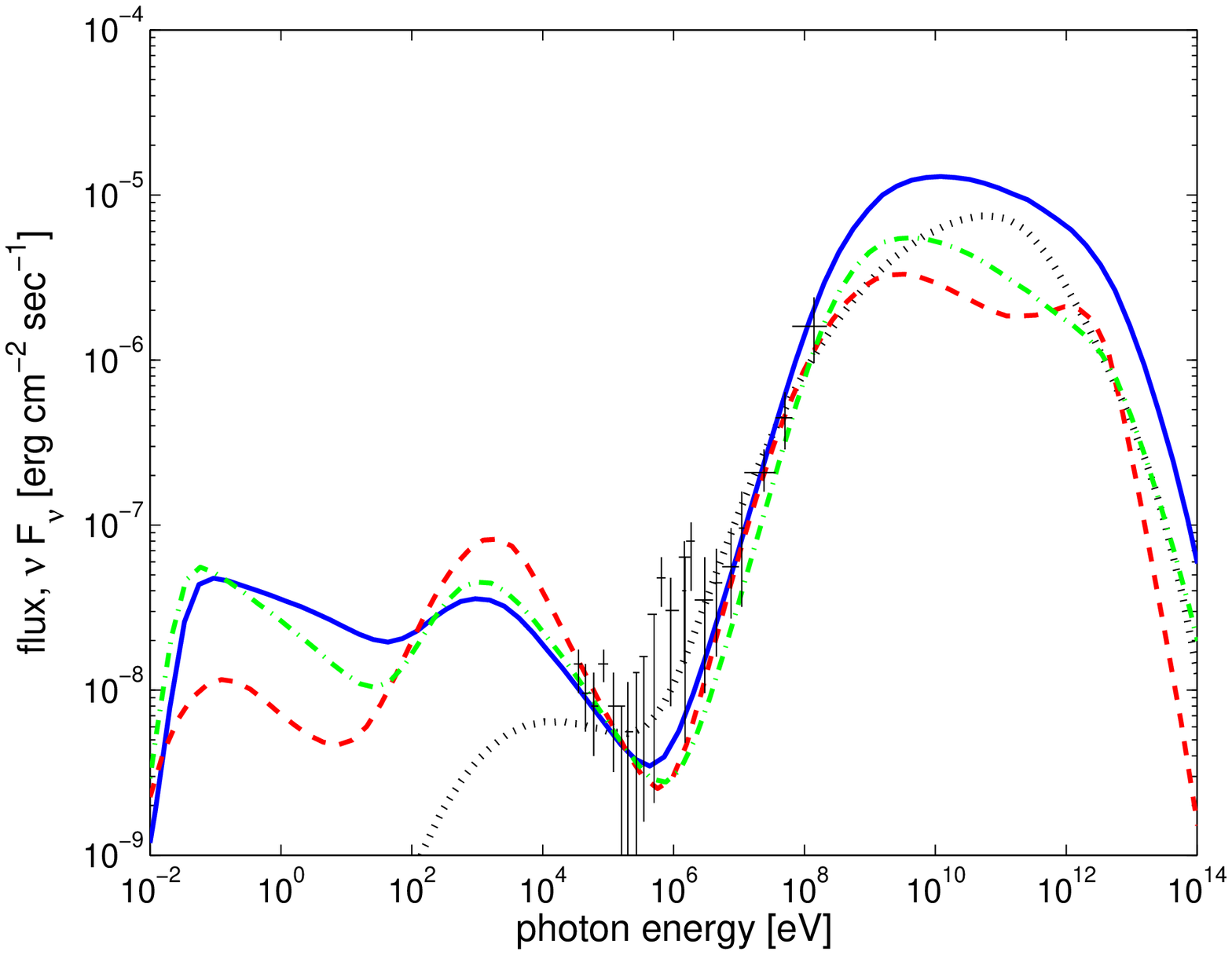}
\caption{
(a) The SSC emission from the forward shock region in the deceleration phase. Temporal evolution of
the theoretical models for synchrotron and SSC components for $\epsilon_e=0.5$, $\epsilon_B=0.01$; solid curves from top to bottom are at onset, 1 min, 1 hour, 1 day, 1 month. The contributions to
the emission at onset are shown as long-dashed (electron-synchrotron),
short-dashed (proton-synchrotron) and dotted (electron IC) curves \cite{ZhangMeszaros01}. (b) Fit to the prompt emission data of GRB 941017 using
the 
IC model of Ref.~\cite{PeerWaxman04a}.}
\label{SSC}
\end{figure*}
A GRB fireball would be significantly decelerated by the circumburst medium starting 
from a distance of $10^{16}-10^{17}$ cm from the central 
engine, at which point
a pair of shocks propagate into the circumburst
medium and the ejecta, respectively. 
Both shocks contain a similar amount of energy. Electrons from 
either shock region
would Compton scatter the soft seed synchrotron
photons from both regions 
to
produce high energy photons
\cite{MeszarosRees94,GranotGuetta03a,PeerWaxman04a,ZhangMeszaros01,Dermeretal00,Wangetal01a,Wangetal01b}.
Compared with the internal shock radius, the deceleration radius corresponds to 
a low ``compactness'' so that high energy
photons 
more readily
escape from the source. 
Figure~\ref{SSC}(a) presents the theoretical forward shock high energy emission components as a function of time for the regime of IC dominance (from \cite{ZhangMeszaros01}).
It is evident that during the first several minutes of the deceleration time, the high energy emission could extend to beyond $\sim$ 10 TeV.
Detection of this emission by ground-based VHE detectors, for sources close enough to have little
absorption by the IR background, would be an important test of this paradigm.


Various IC processes have been considered to interpret the distinct high energy
component detected in GRB 941017 \cite{Gonzalezetal03,GranotGuetta03a}. 
For preferable parameters, the IC emission of forward shock electrons off the
self-absorbed reverse shock emission can interpret the observed spectrum 
(Fig.~\ref{SSC}b, \cite{PeerWaxman04a}). 

\subsection{Steep decay}

Swift observations revealed new features of the GRB afterglow. A canonical X-ray
lightcurve generally consists of five components \cite{Zhangetal06,Nouseketal06}: a steep decay component (with decay index $\sim -3$ or 
steeper), a shallow decay component (with decay index $\sim -0.5$ but
with a wide variation), a normal decay component (with decay index 
$\sim -1.2$), a putative post-jet-break component seen in a small group
of GRBs at later times, and multiple X-ray flares with sharp rise and 
decay occurring 
in nearly half GRBs. Not all five components appear in
every GRB, and the detailed afterglow measurements of GRB 080319B
\cite{Racusin08}  present some challenges to the standard picture we describe here.
The steep decay component \cite{Tagliaferrietal05} is 
generally interpreted as the tail of the prompt gamma-ray emission 
\cite{Zhangetal06,Nouseketal06,KumarPanaitescu00}. 
Within this interpretation,
the steep decay phase corresponds to significant reduction of high energy
flux as well. On the other hand, Ref.~\cite{Dermer07} suggests that the steep decay
is the phase when the blastwave undergoes a strong discharge of its hadronic
energy. Within such a scenario, strong high energy emission 
of hadronic origin is expected. Detection/non-detection of strong high energy
emission during the X-ray steep decay phase would greatly constrain the origin
of the steep decay phase.

\subsection{Shallow decay}

The shallow decay phase following the steep decay phase is still not
well understood \cite{Meszaros06,Granot06,Zhang07}. The standard 
interpretation is that the external forward
shock 
is continuously refreshed by late energy injection, either
from a long-term central engine, or from slower shells ejected in the prompt
phase \cite{Zhangetal06,Nouseketal06,Panaitescuetal06a,GranotKumar06}.
Other options include delay of transfer of the fireball energy to the
medium \cite{KobayashiZhang07}, a line of sight outside the region of prominent
afterglow emission \cite{EichlerGranot06}, a two-component jet model
\cite{Granotetal06}, and time varying shock micro-physics
parameters \cite{Granotetal06,Iokaetal06,Panaitescuetal06b}.

Since the pre-Swift knowledge of the 
afterglow kinetic energy comes from the late afterglow observations, the
existence of the shallow decay phase suggests that the previously estimated
external SSC emission strength is over-estimated during the early afterglow. A modified SSC model 
including the energy injection effect indeed gives less significant SSC
flux 
\cite{GouMeszaros07,Fanetal07}. 
The SSC component nonetheless is still 
detectable by Fermi and higher energy detectors 
for some choices of parameters.  Hence, detections or limits from VHE 
observations constrain those parameters.
If, however, the shallow decay
phase is not the result of a smaller energy in the afterglow shock at
early times, compared to later times, but instead due to a lower
efficiency in producing the X-ray luminosity, the luminosity at higher
photon energies could still be high, and perhaps comparable to (or even
in excess of) pre-Swift expectations. Furthermore, the different
explanations for the flat decay phase predict different high-energy
emission, so 
the latter could help distinguish between the various
models. For example, in the energy injection scenario, the reverse shock
is highly relativistic for a continuous long-lived relativistic wind from
the central source, but only mildly relativistic for an 
outflow that was
ejected during the prompt GRB with a wide range of Lorentz factors and that
gradually catches up with the afterglow shock. 
The
different expectations for the high-energy emission 
in these two cases
may be tested against
future observations.

\subsection{High-energy photons associated with X-ray flares}

X-ray flares have been detected during the early afterglows in a
significant fraction of gamma-ray bursts (e.g. 
\cite{Burrowsetal05,Chincarinietal07,Falconeetal07}). The amplitude of an X-ray flare with 
respect to the background afterglow flux can be up to
a factor of $\sim$500 and the fluence can approximately equal 
the 
prompt emission fluence (e.g. GRB 050502B \cite{Burrowsetal05,Falconeetal06}). The rapid rise
and decay behavior of some flares suggests that they are caused by
internal dissipation of energy due to late central engine
activity \cite{Zhangetal06,Burrowsetal05,Falconeetal06,Liangetal06}. 
There are two likely processes that can produce very
high energy (VHE) photons. One process is that the inner flare
photons, when passing through the forward shocks, would
interact with the shocked electrons and get boosted to higher
energies. Another process is the SSC 
scattering within the X-ray flare region \cite{Fanetal07,Wangetal06}.

\begin{figure*}[!th]
\centering
\includegraphics[width=5.9in]{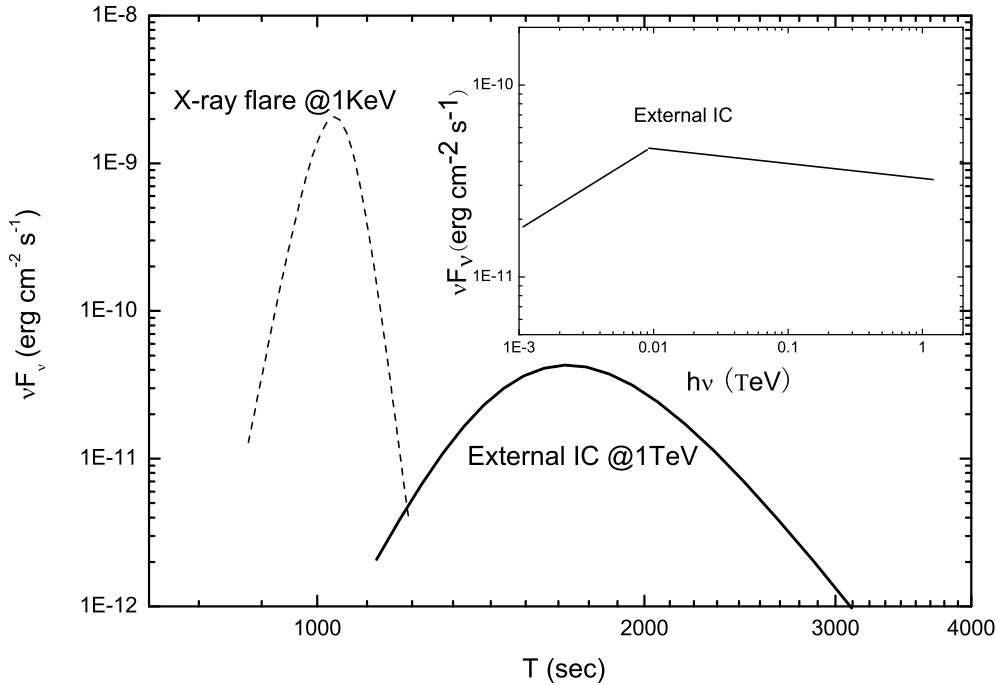}
\caption{The expected light
curves (main figure) and spectral energy distribution (insert
figure) of 
IC scattering of 
X-ray
flare photons by 
forward shock electrons. The flux is
calculated according to 
\cite{Wangetal06}, based
on the following parameters: 
$10^{53}$ erg blast wave energy, 
electron energy distribution index $2.2$, 
electron equipartition factor $\epsilon_e=0.1$, 
1 keV peak energy of the X-ray flare,
$10^{28.5}$ cm source distance
and that the flare has $\delta t/t_f=0.3$. }
\label{flare}
\end{figure*}

Figure~\ref{flare} shows an example of IC scattering of flare 
photons by the afterglow electrons for a flare of duration
$\delta t$ superimposed upon an underlying power law X-ray
afterglow around time $t_f=1000 \,{\rm s}$ after the burst, as
observed in GRB 050502B. 
The duration of the IC emission 
is lengthened by the angular spreading effect
and the anisotropic scattering effect as well 
\cite{Fanetal07,Wangetal06}.   
Using the calculation of \cite{Wangetal06},
for typical parameters as given in the caption,
$\nu F_\nu$ at 1 TeV reaches about $4\times10^{-11}$ erg cm$^{-2}$ s$^{-1}$, with a total duration of about 2000 s.

The
peak energy of the 
SSC scattering within the X-ray flare region
lies at tens of
MeV \cite{Wangetal06} to a few hundreds of MeV \cite{Fanetal07}.
The 
flares may come from internal dissipation processes
similar to the prompt emission, so their dissipation radius may be
much smaller than that of the afterglow external shock. A smaller
dissipation radius causes strong internal absorption to very high
energy photons. For a flare with luminosity $L_x\sim10^{48}$ 
erg s$^{-1}$ and duration $\delta t=100$ s, 
the VHE photons can escape 
only if the dissipation radius is
larger than $\sim$10$^{16}$ cm.  
So in
general, even for a strong X-ray flare occurring at small
dissipation radius, the SSC 
emission at TeV
energies should be lower than the 
IC component above.

\subsection{High-energy photons from\\external reprocessing}

Very high energy photons above 100 GeV produced by GRBs at
cosmological distances are subject to photon-photon attenuation 
by the 
CIB 
(
e.g.\ \cite{Primacketal99,Steckeretal06}) and 
CMB. 
The attenuation of $E$ 
TeV photons by the CIB would produce
secondary electron-positron pairs with a Lorentz factor of
$\gamma_e\simeq 10^6 E$, 
which in turn IC scatter off CMB photons to
produce  MeV--GeV emission 
\cite{Razzaqueetal04,ChengCheng96,DaiLu02,Wangetal04}. This emission is 
delayed relative to the primary
photons 
by two
mechanisms: one is the 
opening angles of
the scattering processes, producing a 
deviation from the direction of the original TeV photons by
an angle $1/\gamma_e$;  the other 
is 
the deflection of 
the secondary pairs in
the intergalactic magnetic field \cite{Plaga95}. Only if the intergalactic
magnetic field is 
less than $\sim$10$^{-16}$ G would the
delayed secondary gamma-rays still be beamed from the same direction
as the GRB.  

\section{High Energy Emission Predictions for Short Bursts}

Recent observational breakthroughs 
\cite{Gehrelsetal05,Foxetal05,Barthelmyetal05,Bergeretal05b,Bloometal06} suggest that at least some short GRBs 
are nearby low-luminosity GRBs that are associated with old stellar populations and 
likely to be 
compact star mergers. The X-ray afterglows of short duration
GRBs are typically much fainter than those of long GRBs, which is consistent with
having a smaller total energy budget and a lower density 
environment 
as expected from
the compact star merger scenarios. Observations suggest
that except being fainter, the afterglows of short GRBs are not distinctly
different from those of long GRBs. 
The long duration
GRB 060614 has a short, hard emission episode followed by extended softer 
emission. It is 
a nearby GRB, but 
has no supernova association, suggesting 
that 060614-like GRBs are 
more energetic versions of short
GRBs \cite{Gehrelsetal06,Zhangetal07}. 

The radiation physics of short GRBs is believed to be similar to that of
long GRBs. As a result, all the processes discussed
above for long GRBs are relevant to short GRBs as well. 
The predicted prompt emission spectrum of a bright short GRB is presented in
Fig.~\ref{short}b 
\cite{GuptaZhang07}. 
Figure~\ref{short}b
is calculated for 
a comparatively bright, 1-second burst at redshift 0.1
with isotropic-equivalent luminosity 10$^{51}$ erg s$^{-1}$. 
Fig.~\ref{short}b 
suggests that 
the high energy component of such a burst is barely detectable by Fermi. Due to 
internal optical depth, the spectrum is cut off beyond about 100 GeV.  VHE observations can constrain the bulk Lorentz factor, since VHE 
emission can be 
achievable if the bulk
Lorentz factor is even larger (e.g. 1000 or above).


No evidence of strong reverse shock emission from short GRBs exists.
For the forward shock, the flux is typically nearly 100 times fainter than
that of long GRBs. This is a combination of low isotropic energy and presumably
a low ambient density. The SSC component in the forward shock region still
leads to GeV-TeV emission, but the flux 
is scaled down by the same factor
as the low energy afterglows. 
Multiple late-time X-ray flares have been detected
for some short GRBs (e.g. GRB 070724 and GRB 050724),
with at least some properties similar to the flares in long GRBs,
so that the emission mechanisms discussed above for long GRB flares
may 
also apply, 
scaled down accordingly. 
In general, short GRBs
may be
less prominent emitters of high energy photons than long GRBs, mainly due
to their low fluence observed in both prompt emission and afterglows. A potential
higher bulk Lorentz factor on the other hand facilitates the escaping of 100 GeV
or even TeV photons from the internal emission region. Furthermore, a few short GRBs are detected at redshifts lower than 0.3, 
and the average short GRB redshift is much lower than that of long GRBs. This is favorable for TeV detection since
the CIB 
absorption 
is greatly reduced at these
redshifts.



\section{Supernova-associated gamma-ray bursts}

Nearby GRBs have been 
associated with
spectroscopically identified supernovae, {\it e.g.}, 
GRB 980425/SN 1998bw, GRB 031203/SN 2003lw, GRB 060218/SN 1006aj, and
GRB 030329/SN 2003lw. 
The processes discussed in the section on high-energy emission from
long GRBs can 
all apply in 
these 
bursts, and
with 
their  close distances, 
VHE emission from 
these sources
would not be significantly attenuated by the CIB{}. 
These bursts
have low luminosity, but the internal absorption by 
soft prompt
emission photons may therefore be lower, so that VHE photons originating from the
internal shock are more likely to escape without significant
absorption, compensating for the overall low flux.
In addition, if there is a highly relativistic jet component
associated with the supernovae, 
supernova
shock breakout photons would be scattered to high energies by the
shock-accelerated electrons in the forward shocks \cite{WangMeszaros06}. 
The strong thermal X-ray emission
from GRB 060218 
may be such 
a relativistic supernova shock 
breakout \cite{Campanaetal06,Waxmanetal07}. 
It has been shown \cite{WangMeszaros06} 
that if the wind mass loss rate from
the progenitor star is 
low, the $\gamma\gamma$ absorption
cutoff energy at early times can be larger than $\sim$100 
GeV, so VHE emission could be detected from these nearby
SN-GRBs.

\section{Ultra High Energy Cosmic Rays and GRBs}

The origin of the UHE cosmic rays (UHECR) is an important
unsolved problem. The idea that they originate from 
long duration GRBs
is argued for a number of reasons. 
%
First, 
the power required for the cosmic rays above the ``ankle'' ($\sim
10^{19}$~eV) is within one or two orders of magnitude equal to the
hard X-ray/$\gamma$-ray power of BATSE GRBs, assumed to be at average
redshift unity \cite{Waxman95,Vietri95,Dermer02}. 
Second, GRBs form powerful relativistic flows, providing extreme sites for
particle acceleration consistent with the known physical limitations, e.g. size,
required to achieve ultra high energy. 
Third, GRBs are expected to be associated with star-forming galaxies, so
numerous UHECR sources would be found within the $\sim 100$~Mpc GZK
radius, thus avoiding the situation that there is no persistent
powerful source within this radius.  
And, finally,  
various features in the medium- and high-energy $\gamma$-ray spectra of
GRBs may 
be attributed to hadronic emission processes.

The required Lorentz factors of UHECRs, $\gtrsim 10^{10}$, exceed by
orders of magnitude the baryon-loading parameter $\eta \gtrsim 100$
thought typical of GRB outflows. Thus the UHECRs must be accelerated
by processes in the relativistic flows. The best-studied mechanism is
Fermi acceleration at shocks, including external shocks when the GRB
blast wave interacts with the surrounding medium, and internal shocks
formed in an intermittent relativistic wind. 

Protons and ions with nuclear charge $Z$ are expected to be
accelerated at shocks, just like electrons.  
The maximum energy in the internal shock model~\cite{Waxman95} 
or in the case of an external shock in a uniform density medium~\cite{Vietri98,DermerHumi01} are both of order a few $Z$ 10$^{20}$ eV for
typical expected burst parameters. 
Thus GRBs can 
accelerate UHECRs.  The ultrarelativstic
protons/ions in the GRB jet and blast wave can interact with ambient
soft photons 
if the corresponding opacity
is of the order of unity or higher, to form escaping neutral
radiations (neutrons, $\gamma$-rays, and neutrinos).  They may also
interact with other baryons via inelastic nuclear production
processes, again producing neutrals.  
So VHE gamma rays are a natural consequence of UHECR acceleration in 
GRBs.
While leptonic models explain keV--MeV data as synchrotron or
Compton radiation from accelerated primary electrons, and GeV--TeV
emission from inverse-Compton scattering, a hadronic emission component
at GeV--TeV energies can also be present.



Neutrons are coupled to the jet protons by elastic
$p$-$n$ nuclear scattering and, depending on injection conditions in the
GRB, can decouple from the protons during the expansion phase.  As a
result, the neutrons and protons travel with different speeds and will
undergo inelastic $p$-$n$ collisions, leading to $\pi$-decay radiation,
resulting in tens of GeV photons~\cite{Derishevetal99,BahcallMeszaros00}.  The decoupling
leads to subsequent interactions of the proton and neutron-decay
shells, which may reduce the shell Lorentz factor by heating~\cite{Rossietal06}.
The $n$-$p$ decoupling occurs in short GRBs for values of the
baryon-loading parameter $\eta \sim 300$ \cite{RazzaqueMeszaros06a}.  The relative
Lorentz factor between the proton and neutron components may be larger
than in long duration GRBs, leading to energetic ($\sim$50 GeV)
photon emission. 
Applying this model to several short GRBs in the field of view of 
Milagro~\cite{Abdoetal07} gives fluxes of a few 10$^{-7}$ cm$^{-2}$ s$^{-1}$ for
typical bursts, suggesting that a detector of large
effective area, $\gtrsim 10^7$~cm$^{2}$, at low threshold energy is
needed to detect these photons.  For the possibly nearby ($z=0.001$) GRB 051103,
the flux could be as large as $\sim$10$^{-3}$ cm$^{-2}$ s$^{-1}$.

%


Nuclei accelerated in the GRB jet and blast wave to
ultra high energies can make $\gamma$-rays through the synchrotron
process; 
photopair production, which converts the target photon
into an electron-positron pair with about the same Lorentz factor
as the ultrarelativistic 
nucleus; and 
photopion production, which makes pions that decay into 
electrons and positrons, photons, and neutrinos.
The target photons for the 
latter two
processes are usually considered
to be the ambient synchrotron and synchrotron self-Compton photons
formed by leptons accelerated at the forward and reverse shocks of
internal and external shocks.  If the pion-decay muons decay before
radiating much energy~\cite{RachenMeszaros98}, the secondary leptons,
$\gamma$-rays, and neutrinos each carry about 5\% of the primary
energy.

About one-half of the time, neutrons are formed in a photopion
reaction.  If the neutron does not undergo another photopion reaction
before escaping the blast wave, it becomes free to travel until it
decays. Neutrons in the neutral beam \cite{AtoyanDermer03}, collimated by the
bulk relativistic motion of the GRB blast wave shell, travel $\approx
(E_n/10^{20}$ eV) Mpc before decaying. A neutron decays into neutrinos
and electrons with $\approx 0.1$\% of the energy of the primary.
Ultrarelativistic neutrons can also form secondary pions after
interacting with other soft photons in the GRB enviroment. The
resulting decay electrons form a hyper-relativistic synchrotron
spectrum, which is proposed as the explanation for the anomalous
$\gamma$-ray emission signatures seen in GRB 941017 \cite{DermerAtoyan04}.

The electromagnetic secondaries generate an electromagnetic cascade
when the optical depth 
is sufficiently large. The photon number
index of the escaping $\gamma$-rays formed by multiple generations of
Compton and synchrotron radiation is generally between $-3/2$ and $-2$
below an exponential cutoff energy, which could reach to GeV or,
depending on parameter choices, TeV energies 
\cite{DermerAtoyan03,AtoyanDermer03}.


Gamma-ray observations of GRBs will help distinguish between leptonic
and hadronic emissions.  
VHE 
$\gamma$-ray emission
from GRBs can be modeled by synchro-Compton processes of
shock-accelerated electrons 
\cite{PeerWaxman04b,Meszarosetal94,PapathanassiouMeszaros96,ChiangDermer99,Dermeretal00}, or by
photohadronic interactions of UHECRs and subsequent cascade emission
\cite{BottcherDermer98,Totani99,Fragileetal04}, or by a combined leptonic/hadronic model.
The clear distinction between the two models from $\gamma$-ray
observation will not be easy. The fact that the VHE $\gamma$-rays are
attenuated both at their production sites and in the 
CIB
restricts measurements to energies below  150 GeV ($z \sim 1$) or 5
TeV ($z \lesssim 0.2$).  Distinctive features of 
hadronic models are:
\begin{itemize}
\item 
Photohadronic interactions and subsequent electromagnetic $\gamma$-ray
producing cascades develop over a long time scale due to slower energy
loss-rate by protons than electrons. The GeV-TeV light curves arising
from hadronic mechanisms then would be longer than those expected from
purely leptonic processes \cite{BottcherDermer98},
facilitating detection with pointed instruments. 
\item
Cascade $\gamma$ rays will be harder than a $-2$ spectrum below an
exponential cutoff energy, and photohadronic processes can make hard,
$\sim-1$ spectra from anisotropic photohadronic-induced cascades,
used to explain GRB 941017 \cite{Gonzalezetal03,DermerAtoyan04}. A ``two zone'' leptonic
synchro-Compton mechanism can, however, also explain the same
observations \cite{GranotGuetta03a,PeerWaxman04a,ZhangMeszaros01,Wangetal05}, with low energy emission
from the prompt phase and high energy emission from a very early
afterglow.
\item
Another temporal signature of hadronic models is delayed emission from 
UHECR cascades in the CIB/CMB \cite{WaxmanCoppi96} or
$\gtrsim$PeV energy $\gamma$-rays, from $\pi^0$ decay, 
which may escape the GRB fireball \cite{Razzaqueetal04}. 
However, $\gtrsim$TeV
photons created by leptonic synchro-Compton mechanism in external
forward shocks may imitate the same time delay by cascading in the
background fields \cite{Wangetal04}.
\item 
Quasi-monoenergetic $\pi^0$ decay $\gamma$-rays from $n$-$p$
decoupling, which are emitted from the jet photosphere prior to the
GRB, is a promising hadronic signature \cite{Derishevetal99,BahcallMeszaros00,RazzaqueMeszaros06a}, though
it requires that the GRB jet should contain abundant free neutrons as
well as a large baryon load.
\end{itemize}
Detection of high-energy neutrino emissions would conclusively
demonstrate 
cosmic ray 
acceleration in GRBs,
but non-detection would not conclusively rule out GRBs as a source
of UHECRs, since the $\nu$ production level even for optimistic parameters
is small.
\section{Tests of Lorentz Invariance with Bursts}

Due to quantum gravity effects, it is possible that the speed of light is energy dependent and that ${\Delta}c/c$ scales either linearly or quadratically with ${\Delta}E/E_{QG}$, where $E_{QG}$ could be assumed to be at or below the Planck energy, $E_{P}$  
\cite{ellis1992,amelino1997,gambini1999}. Recent detections of flaring from the blazar Mrk 501, using the MAGIC IACT, have used this effect to constrain the quantum gravity scale for linear variations to $\gtrsim 0.1E_{P}$ \cite{albert2007}. This same technique could be applied to GRBs, which have fast variability, if they were detected in the TeV range and if the intrinsic chromatic variations were known.
However, there may be intrinsic limitations 
to some approaches \cite{Scargleetal07}.
By improving the sensitivity and the energy range with a future telescope array, the current limit could be more tightly constrained, particularly if it were combined with an instrument such as Fermi at lower energies, thus increasing the energy lever arm.

\begin{figure*}[!th]
\centering
\includegraphics[width=5in]{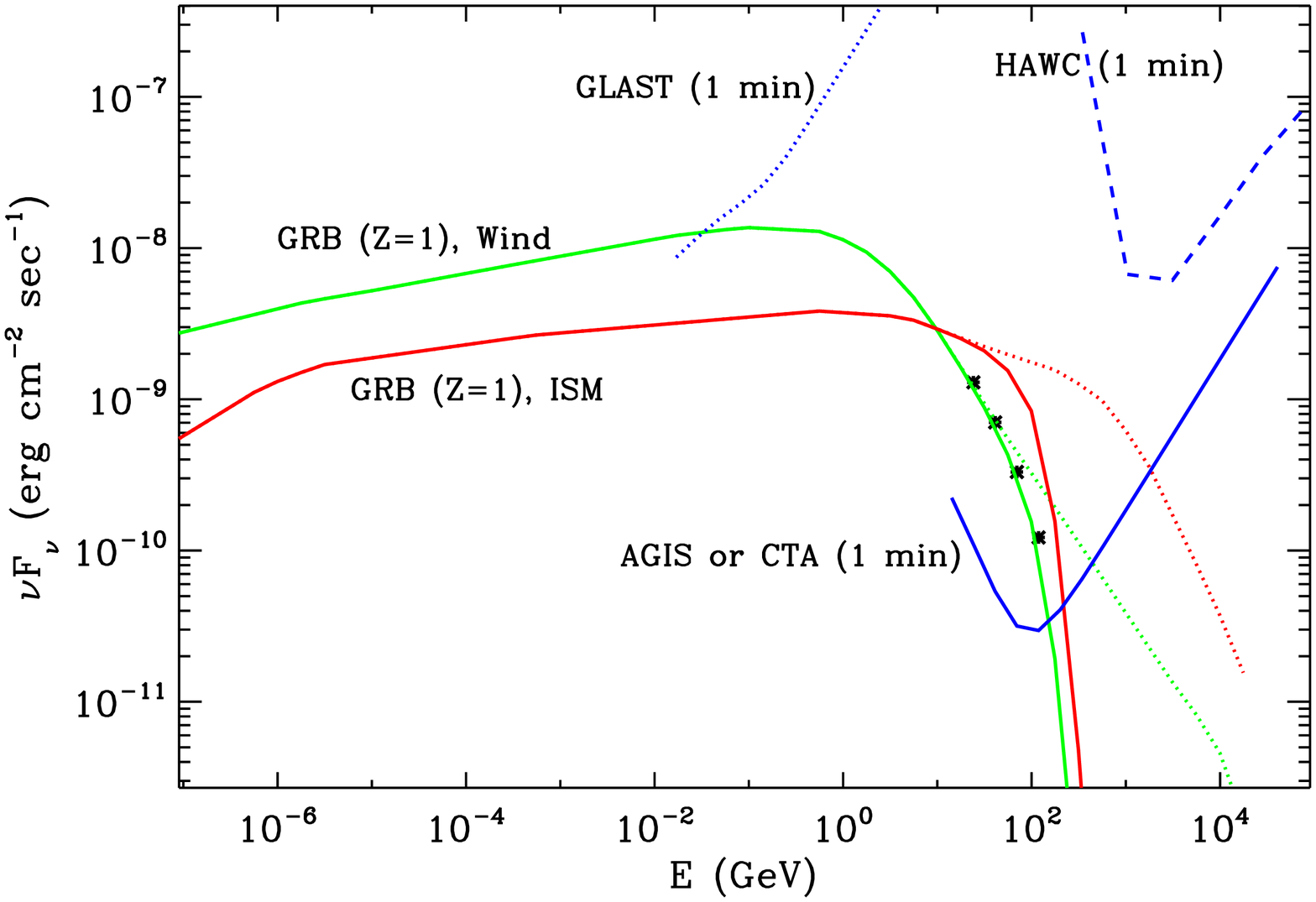}
\caption{A plot of the predicted gamma-ray spectrum from a GRB at a redshift of z=1 adapted from Pe'er and Waxman 
\cite{PeerWaxman05b},
reduced by a factor of 10 to illustrate the sensitivity even to 
weaker bursts.
The green and red curves show the calculation for a wind environment and an ISM-like environment. The dotted curves give the source spectrum, while the solid curves include the effects of intergalactic absorption using a model from Franceschini et al.\  \cite{Franceschinietal2008}.
The blue curves show the differential sensitivity curves for Fermi (GLAST; dotted), a km$^2$ IACT array like AGIS or CTA (solid) and 
the HAWC air shower array (dashed). 
For the AGIS/CTA curve we show the differential sensitivity for 
0.25 decade bins, while for the HAWC instrument we assume 
0.5 decade bins.
The sensitivity curve is based on a 5 sigma detection and at least 25 detected photons. Black points and error bars (not visible) are simulated independent spectral points that could be obtained with AGIS/CTA.}
\label{grb_sedsim}
\end{figure*}

\section{Detection Strategies for VHE Gamma-Ray Burst Emission}



Ground-based observations of TeV emission from gamma ray bursts are difficult. The fraction of GRBs close enough to 
elude attenuation 
at TeV energies by the CIB 
is small. Only $\sim$10\% of long bursts are within z$<$0.5, 
the redshift of the most distant detected VHE 
source, 3C 279 \cite{Teshima:07}. Short bursts are more nearby with over 50\% detected within z$<$0.5, but the prompt emission has ended prior to satellite notifications of the burst location.

Therefore, wide field of view detectors with high duty cycle operations 
would be ideal to observe the prompt emission from 
gamma-ray bursts. Imaging atmospheric Cherenkov telescopes (IACT) can be made 
to cover large sections of the sky by either having many 
mirrors each pointing in a separate direction or by employing secondary optics to expand the field of view of each mirror. 
However, the duty factor is still $\sim$10\% 
due to solar, lunar, and weather constraints. IACTs could also be made with fast slewing mounts to allow them to slew to most GRBs within $\sim$20 seconds, thus allowing them to observe some GRBs before the end of the prompt phase. 
Alternatively, extensive air shower 
detectors intrinsically have a field of view of $\sim$2 sr and operate with $\sim$95\% 
duty factor. These observatories, especially if located at very high altitudes, can detect gamma rays down to 100 GeV, but at these low energies they lack good energy resolution and have a point spread function of $\sim$1 degree. 
The traditionally less sensitive extensive air shower detectors may have difficulty achieving the required prompt emission sensitivity on short timescales ($>5\sigma$ detection of 10$^{-9}$ erg s$^{-1}$ cm$^{-2}$ in $\lesssim20$ sec integration). The combined observations of both of these types of detectors would yield the most complete picture of the prompt high energy emission. 
The expected performance of the two techniques relative to 
a particular 
prompt GRB emission 
model 
is shown in Figure~\ref{grb_sedsim}. 

The detector strategy for extended emission associated with traditional afterglows or with late-time flares from GRBs is far simpler than the strategy for early prompt emission. The high sensitivity and low energy threshold of an IACT array are the best way to capture photons from this emission at times greater than $\sim$1 min, particularly if fast slewing is included in the design.
\section{Synergy with other instruments}
While GRB triggers are possible from wide angle VHE instruments, a space-based GRB 
detector will be needed. Swift, Fermi, or future wide field of view X-ray monitors such as EXIST or JANUS must provide lower energy observations. GRBs with observations by both Fermi and VHE 
telescopes will be particularly exciting and may probe high Lorentz factors. Neutrino 
telescopes such as IceCube, UHECR telescopes such as Auger, and next generation VHE observatories 
can supplement one another in the search for UHECRs from GRBs, since neutrinos are 
expected along with VHE gamma rays. 
Detection of gravitational waves from GRB progenitors with instruments such as LIGO have the potential to reveal the engine powering the GRB fireball.  Correlated observations between gravitational wave observatories and VHE gamma-ray instruments will then 
be important for understanding which type(s) of engine can power
VHE emission.

Correlated observations 
between TeV gamma-ray detectors and neutrino detectors 
have 
the potential for significant reduction in background for the
participants. 
If TeV gamma sources are observed, observers will know where and when to look for neutrinos (and vice versa 
\cite{KowalskiMohr07}, 
though the advantage in that direction is less significant).  
For example, searches for GRB neutrinos have used the known time and
location to reduce the background
by a factor of nearly 10$^5$ compared to an annual all-sky diffuse search \cite{Achterbergetal07}.
Beyond decreasing background, correlated observations also have the potential to increase the expected signal rate.  
If the spectrum of high-energy gamma rays is known, then constraints on the expected neutrino spectrum can also be introduced, allowing the signal-to-noise ratio of neutrino searches to be 
significantly improved \cite{StamatikosBand06}.
In the case of the AMANDA GRB neutrino search, which is based on a specific 
theoretical neutrino spectrum, the expected signal collection efficiency is nearly
20 times higher than the less constrained search for diffuse UHE
neutrinos.
With combined photon and neutrino observational efforts, there is a much better chance of eventual neutrino detection of sources such as GRBs (and AGN).



\section{Conclusions}




Gamma-ray bursts undoubtedly involve a population of high-energy 
particles responsible for the emission detected
from all bursts (by definition) at energies up to of order 1 MeV,
and for a few bursts so far observable by EGRET, up to a few GeV.  Gamma-ray 
bursts may in fact be the source of the highest energy particles in the universe.
In virtually all models, this high-energy population can also produce
VHE gamma-rays, although in many cases the burst environment would be 
optically thick to their escape.  The search for and study of VHE emission from GRB therefore tests theories about the nature of these
high energy particles (Are they electrons or protons? What is their
spectrum?) and their environment (What are the density and bulk
Lorentz factors of the material? What are the radiation fields?
What is the distance of the emission site from the central source?).
In addition, sensitive VHE measurements would aid in assessing the 
the total calorimetric radiation output from bursts.  
Knowledge of the VHE gamma-ray properties of bursts will therefore
help complete the picture of these most powerful known accelerators.


An example of the insight that can be gleaned from VHE data
is that leptonic synchrotron/SSC
models can be tested, and model parameters extracted, by 
correlating the peak energy of
X-ray/soft $\gamma$-ray emission
with GeV--TeV data.
For long lived GRBs, the spectral properties of late-time flaring in the X-ray band can be compared to the measurements in the VHE band, where associated emission is expected.
Of clear interest is whether there are distinctly evolving high-energy $\gamma$-ray  
spectral components, whether at MeV, GeV or TeV energies, unaccompanied
by the 
associated lower-energy component 
expected in leptonic
synchro-Compton models. 
Emission of this sort is most easily explained in models involving
proton acceleration.
As a final example, the escape of VHE photons from the burst fireball provides a tracer of the minimum Doppler boost and bulk Lorentz motion of the emission region 
along the line of site, since the inferred opacity of the emission region declines with increasing boost.   






There are observational challenges for detecting VHE emission during the initial
prompt phase of the burst.  The short duration of
emission leaves little time (tens of seconds) for repointing 
an instrument, and the opacity of the compact fireball is at its 
highest.  For the majority of bursts having redshift $\gtrsim$0.5, the 
absorption of gamma rays during all phases of the burst by 
collisions with the extragalactic background light reduces the 
detectable emission, more severely with increasing gamma-ray energy.
With sufficient sensitivity, an all-sky
instrument is the most desirable for studying the prompt phase,
in order to measure the largest sample of bursts and to catch 
them at the earliest times.
As discussed in the report of the Technology Working Group,
the techniques used to implement all-sky compared to 
pointed VHE instruments result in a trade-off of energy threshold and
instantaneous sensitivity for field of view.  More than an order of magnitude improvement in sensitivity to GRBs is envisioned
for the next-generation instruments of both types, giving both 
approaches a role in future studies of GRB prompt emission.

The detection of VHE afterglow emission, delayed prompt emission from large radii, and/or late X-ray flare-associated emission simply requires a sensitive instrument 
with only moderate slew speed. It is likely that an instrument with significant sensitivity improvements over the current generation of IACTs will detect GRB-related VHE emission from one or all of these mechanisms which do not suffer from high internal absorption, thus making great strides towards understanding the extreme nature and environments of GRBs and their ability to accelerate particles.

In conclusion, large steps in understanding GRBs have frequently resulted from particular new characteristics measured for the first time in a single burst. 
New instruments improving sensitivity to very-high-energy gamma-rays
by an order or magnitude or more compared to existing observations
have the promise to make just
such a breakthrough in the VHE band.

\end{document}